\documentclass{SCGE}

\begin{document}

\begin{picture}(0,0){\rm
\put(0,-20){\makebox[160truemm][l]{\bf {\sanhao\raisebox{2pt}{.}}
Article  {\sanhao\raisebox{1.5pt}{.}}}}}
\put(0,-34){\jiuwuhao {\textcolor[rgb]{0.5,0.5,0.5}{\sf 
}}}
\end{picture}

\def\bm{\boldsymbol}

\def\dl{\displaystyle}
\def\du{\end{document}}
\def\d{{\rm d}}
\def\e{{\rm e}}
\def\i{{\rm i}}

\Year{2014} %
\Month{November} %
\Vol{57} 
\No{11} 
\BeginPage{1} 
\AuthorMark{{\rm Fan X H}, et al.}  
\AuthorMarkCite{\rm Fan X H, Dong J M, Zuo W. } 
\DOI{111} 
\ArtNo{112110}

\title[Symmetry energy at subsaturation densities and the neutron skin thickness of $^{208}\text{Pb}$ ]{Symmetry energy at subsaturation densities and the neutron skin thickness of $^{208}\text{Pb}$}

\author[1,2]{FAN XiaoHua}{}
\author[1]{DONG JianMin}{Corresponding author (email: djm4008@126.com)}
\author[1]{ZUO Wei}{Corresponding author (email: zuowei@impcas.ac.cn)}

\address[{\rm1}]{Institute of Modern Physics, Chinese Academy of Sciences, Lanzhou 730000, China;}
\address[{\rm2}]{University of Chinese Academy of Sciences, Beijing, 100049, China;}
\maketitle \vspace{-3.5mm}{\footnotesize\begin{center} Received Month date, Year; accepted Month date, Year
\end{center}}\vspace*{-5mm}

\begin{center}
\rule{16.5cm}{0.4pt}
\parbox{16.5cm}
{\begin{abstract}The mass-dependent symmetry energy coefficients $a_{sym}(A)$  has been extracted by analysing the heavy nuclear mass differences reducing the uncertainties as far as possible in our previous work. Taking advantage of the obtained symmetry energy coefficient $a_{sym}(A)$ and the density profiles obtained by switching off the Coulomb interaction in $^{208}\text{Pb}$, we calculated the slope parameter $L_{0.11}$ of the symmetry energy at the density of $0.11\text{fm}^{-3}$. The calculated $L_{0.11}$ ranges from 40.5 MeV to 60.3 MeV. The slope parameter $L_{0.11}$ of the symmetry energy at the density of $0.11\text{fm}^{-3}$ is also calculated directly with Skyrme interactions for nuclear matter and is found to have a fine linear relation with the neutron skin thickness of $^{208}\text{Pb}$, which is the difference of the neutron and proton rms radii of the nucleus. With the linear relation the neutron skin thickness $ \Delta R_{np} $ of $^{208}\text{Pb}$ is predicted to be 0.15 - 0.21 fm.
\end{abstract}}
\end{center}\vspace*{-0.6cm}

\begin{center}
\parbox{16.5cm}
{\bf\jiuhao nuclear matter, symmetry energy, neutron skin }
\end{center}

\begin{center}
{}
\CITA    
\end{center}

\textwidth=178truemm \textheight=236truemm

\wuhao\vspace*{1.5mm}

\begin{multicols}{2}

\renewcommand{\baselinestretch}{1.08} \baselineskip 12.2pt\parindent=10.8pt

\renewcommand{\thefootnote}

\section{Introduction}

Symmetry energy is one of the basic features of the equation of state (EOS) of nuclear matter. It represents the energy cost in translating all the protons to neutrons in symmetric nuclear matter for per nucleon approximately. The density-dependent symmetry energy $S(\rho )$ has attracted great attention in nuclear physics and astrophysics such as heavy ion reaction\cite{FRW,AMJ,VMV,BLC,G1,W2,W3,S,F,Y,M,W,G,ZYX}, the stability of superheavy nuclei\cite{DZ}, nuclear structure\cite{Ren1,Ren2}, the structures, composition and cooling of neutron stars \cite{CJ,BS,JM,BJ} and even some new  physics beyond the standard model\cite{CS,TM}. Consequently, the slope and curvature parameters which describe the density-dependence are of much importance to understand a variety of issues in the mentioned areas. The slope parameter $L$ is vital particularly, which mostly decides its density-dependence. At the saturation density, the slope parameter $L$ probed by plenty of methods can vary largely \cite{DZWZ}. Fortunately, recent available constraints on $L$ from terrestrial laboratory measurements and astrophysical observations are in agreement with $L\simeq55\pm25$ MeV, which is summarized by Chen recently \cite{CHEN}. But further efforts are still needed in the determination of the density-dependence of the symmetry energy, in particular, the high density behaviors. Although the nuclear properties given by various effective interactions are very different, the finite nuclei structure characters provided by them are approximately unanimous. For this reason, it is a significant way that we explore the nuclear matter with the help of properties of finite nuclei. In our work, we utilize the properties of finite nuclei to probe the density-dependence of symmetry energy around the subsaturation density of $0.11\text{fm}^{-3}$. \\
\phantom{fff}In recent years, it has been established that the neutron skin thickness $ \Delta R_{np} $  of $^{208}\text{Pb}$ is linearly correlated with the density dependence of the nuclear symmetry energy around saturation\cite{B1,B2,RJ,AW,MXXM,MXX}. In the present work, we also observe that  there is a strong linear relation between the neutron skin thickness $ \Delta R_{np} $ of $^{208}\text{Pb}$ and the slope parameter $L_{0.11}$ of the symmetry energy at the density of $0.11\text{fm}^{-3}$. According to this we can obtain a strong
constraint on the density dependence of $S(\rho)$ with a measurement of $ \Delta R_{np} $ with a high accuracy. Correspondingly, the neutron skin $ \Delta R_{np} $ can be predicted by the the slope parameter $L^{(6)}_{0.11}$. As the smaller neutron skins in heavy nuclei tend to yield smaller neutron star radii \cite{CJJ} given by Horowitz and Piekarewicz and the value of neutron skin play a key role in deciding whether the $1.4M_{\odot}$  neutron stars can have a direct Urca process \cite{AMJ}, there is a great necessariness for an accurate measurement of the neutron skin. The information about the neutron skin thickness of $^{208}\text{Pb}$ $ \Delta R_{np}=0.156^{+0.025}_{-0.021}$ fm from proton inelastic scattering \cite{TA}, $\Delta R_{np}=0.17\pm 0.03$ fm from chiral effective field theory \cite{KMJ} and $ \Delta R_{np}=0.191\pm0.032$ fm from $\alpha$-decay energies \cite{DZG} shows a big challenge of a precise constraint about the neutron skin thickness.      \\
\section{Method}
For heavy nuclei, such as $^{208}\text {Pb}$, the nuclear surface region where nucleon density is much less than the saturation density, contributes dominantly to symmetry energy\cite{DZG}. So it is better to describe the density-dependent symmetry energy by expanding around the density that is below the saturation density.
Around the nuclear matter subsaturation density
$\rho=0.11 \text{fm}^{-3}$, the symmetry energy $S(\rho)$ is expanded to
second order in term of density $\rho$ as

\begin{equation}
S(\rho)=S_{0.11}+\frac{L_{0.11}}{3}\left( \frac{\rho -0.11 }{0.11}\right) +
\frac{K_{\text{sym}}}{18}\left( \frac{\rho-0.11}{0.11}\right)
^{2}.
\end{equation}
There is an available connection that the $a_{sym}(A)$ of finite nuclei is approximately equal to $S(\rho_{A})$ of the nuclear matter at a reference density $\rho_{A}$, which is proposed by Centelles, et al\cite{MXXM}. It makes the symmetry energy of the nuclear
matter in contact with the one of finite nuclei, and thus allows one to explore the density dependence of the symmetry energy $S(\rho)$. We have extracted the mass-dependent symmetry energy coefficients $a_{sym}(A)$  in our previous work and found that  $a_{sym}(A) \simeq 22.4$ MeV for $^{208}$Pb \cite{FD}. The reference density $\rho_{A}$ for $^{208}$Pb is figured out to be $0.55\rho_{0}=0.088$ $ \text{fm}^{-3}$ with $\rho_{0}=0.16$ $  \text{fm}^{-3}$. Substituting the $\rho_{A}$ in Eq.[1], we obtain
\begin{equation}
22.4 ~ \text{MeV}=S_{0.11}-\frac{0.6L_{0.11}-0.02K_{sym}}{9}  .
\end{equation}
There are several shapes of symmetry energy as a function of density\cite{DZJU}. It is found that the expression $S(\rho) = S_{0}(\rho /\rho_{0})^{\gamma}$ or $S(\rho) =12.5 (\rho/\rho_{0})^{2/3} + C_{P} (\rho/\rho_{0})^{\gamma}$ is not universal. But the shape from the density-dependent M3Y interaction\cite{TD} is much better. With it, we can describe the symmetry energy $S(\rho )$ in nuclear matter as the following formulism
\begin{equation}
S(\rho)=C_{K}\left(\frac{\rho}{0.11}\right)^{\frac{2}{3}}
+C_{1}\left(\frac{\rho}{0.11}\right)
+C_{2}\left(\frac{\rho}{0.11}\right)^{\frac{5}{3}},
\end{equation}
where $C_{K}=\frac{\hbar^{2}{K_{F}}^{2}}{6m}\simeq$ 9.6 MeV, $C_{1}$ and $C_{2}$ are two parameters that are required to be determined. $L_{0.11}=3\rho\frac{\partial S}{\partial \rho} |_{0.11}$ and $K_{\text{sym}}=9\rho^{2}\frac{\partial^2 S}{\partial \rho^2}|_{0.11}$ are slope and curvature parameters at the density of 0.11fm$^{-3}$ respectively. There is an useful
connection among the symmetry energy at the density of $0.11 \text{fm}^{-3}S_{0.11}$, $L_{0.11}$ and $K_{\text{sym}}$
\begin{equation}
K_{\text{sym}}=5L_{0.11}-15S_{0.11}+28.8,
\end{equation}
which is sufficient to estimate the small contribution of $K_{\text{sym}}$ in Eq.[1].
As a consequence, Eq.[1] can be rewritten as
\begin{equation}
\begin{aligned}
S(\rho )=&\frac{195.24-0.52L_{0.11}}{9}+\frac{L_{0.11}}{3}\left( \frac{\rho -0.11 }{0.11}\right)\\
&+\frac{4.14L_{0.11}-317.79}{18}\left(\frac{\rho-0.11}{0.11}\right)
^{2}.
\end{aligned}
\end{equation}
Therefore, once the $L_{0.11}$ is determined, the symmetry energy $S(\rho)$ at subsaturation density is obtained. Thus, the centre goal of this work is to determine  $L_{0.11}$.
In the local density approximation, the symmetry energy  coefficient $a_{sym}(A)$ can be calculated \cite{SKM} as
\begin{equation}
a_{\text{sym}}(A)=\frac{1}{AX_{0}^{2}}\int\
d^{3}r\rho(r)S[\rho(r)][\delta (r)]^{2},
\end{equation}
where $X_{0}=(N-Z)/A$ is the whole nuclear isospin asymmetry, $\delta(r)=(\rho_{n}(r)-\rho_{p}(r))/{\rho(r)}$ the isospin asymmetry profile, $\rho(r)$ the whole nucleon density and $a_{\text{sym}}=22.4$ MeV for $^{208}\text{Pb}$. $\rho_{n}(r)$ and $\rho_{p}(r)$ are the density profile of neutron and proton respectively  in nucleus that can be given by the mean field models such as Skyrme-Hartree-Fock.  Here, we calculated $\rho(r)$, $\delta(r)$ and the neutron skin thickness $\Delta R_{np}$ with the Skyrme interactions\cite{MOJA}: SIII, SLy4, SLy5, SkI2, SkI4, SkP, SkM*, SGII, T4, T6, BSk8, LNS1, LNS5, HFB17, KDE0, KDE, SKz2, SKz4, SV, MSk1, MSkA, v090, SK255, SK272, SIV, SkMP, MSL0, SKA, SKSC15 and SKSC40. Substituting Eq.[5] in Eq.[6], the slope parameter $L_{0.11}$ is determined and the obtained $L_{0.11}$ with this method is labeled $L^{(6)}_{0.11}$. Indeed, in order to extract the nuclear symmetry energy that relates solely to the nuclear force, the effect due to the fact that the Coulomb interaction effectively polarizes the neutron and proton densities should be subtracted. So we calculate the neutron density $\rho_{n}(r)$ and proton density  $\rho_{p}(r)$ by switching off the Coulomb interaction. On the contrary, the Coulomb interaction is necessary to work out the neutron skin thickness $\Delta R_{np}$ on account of that the protons and neutrons move slightly away from the core as a result of the Coulomb repulsion between charged protons.

\section{Results}
Known that $a_{\text{sym}}(A)\simeq 22.4$ of heavy spherical nucleus $^{208}\text{Pb}$ above, we can obtain $L_{0.11}$ with various Skyrme interactions. The results given by some typical Skyrme interactions mentioned above are shown in Table 1. $L^{(6)}_{0.11}$ and $S^{(6)}_{0.11}$ are the slope parameter and the symmetry energy at the density of $0.11\text{fm}^{-3}$ calculated with the properties of finite nuclei while $L_{0.11}$ and $S_{0.11}$  are those calculated with the Skyrme interactions for nuclear matter directly. $L_{0.11}$ obtained by the Skyrme procedure for nuclear matter directly has a large scope  but the $L^{(6)}_{0.11}$ computed with Eq.[6] ranges narrowly. And it is the same for the comparison between both symmetry energy, namely $S^{(6)}_{0.11}$ and $S_{0.11}$ gained in the two ways.  It thus appears that the method of calculating the features of nuclear matter by using the structure properties of the finite nuclei is feasible and effective.

\begin{tablehere}
\caption{Comparison between the  values of the slope parameter and the symmetry energy at the density of $0.11\text{fm}^{-3}$  with the properties of finite nuclei namely, Eq.[6], and those calculated with Skyrme interactions directly.} \vspace{1mm}\footnotesize

\setlength{\arrayrulewidth}{0.8pt}
\begin{center} \doublerulesep 0.1pt \tabcolsep 13.5pt
\begin{tabular}{lccccc}
\hline
Force&$L^{(6)}_{0.11}$&$S^{(6)}_{0.11}$&$L_{0.11}$&$S_{0.11}$\\
   &(MeV)&(MeV)&(MeV)&(MeV) \\
 \hline\
SIII&50.71&25.54&32.02&26.06\\
SLy4&48.03&25.34&42.09&26.49\\
SLy5&47.29&25.28&43.52&26.22\\
SkI2&60.30&26.27&68.17&23.23\\
SkI4&49.74&25.47&46.09&22.9\\
SkP&46.33&25.21&35.26&26.21\\
SkM*&50.33&25.51&44.17&24.32\\
SkMP&53.31&25.74&53.45&22.59\\
SGII&45.26&25.13&37.95&22.18\\
T4&59.63&26.22&67.33&25.63\\
T6&52.77&26.70&38.58&25.43\\
BSk8&50.85&25.55&28.82&25.18\\
LNS1&44.54&25.07&38.54&25.31\\
LNS5&45.17&25.12&44.91&23.14\\
HFB17&49.00&25.41&40.08&25.25\\
KDE0&46.42&25.22&43.89&27.30\\
KDE&44.87&25.10&41.05&26.39\\
SKz2&43.53&25.00&33.36&28.69\\
SKz4&43.15&24.97&25.59&29.90\\
SV&42.81&24.94&66.72&23.62\\
MSk1&49.15&25.42&39.69&25.49\\
MSkA&50.07&25.49&50.90&24.31\\
v090&43.35&25.06&27.44&25.84\\
SK255&58.60&26.14&71.00&27.56\\
SK272&58.72&26.15&70.15&28.13\\
SIV&52.04&25.64&55.59&24.91\\
MSL0&52.29&25.66&49.38&23.15\\
SKA&55.55&25.91&58.98&25.25\\
SKSC15&43.15&24.97&27.99&25.58\\
SKSC40&40.51&24.77&21.12&25.98\\

 \hline
\end{tabular}
\end{center}

\end{tablehere}

All the results about $L_{0.11}$  from Eq.[6] (labeled $L^{(6)}_{0.11}$) are shown in the left panel of Fig.1 between the two dashed lines. The calculated $L_{0.11}$ directly for the nuclear matter is shown in the horizontal axis and the the neutron skin thickness $\Delta R_{np}$ is shown in the vertical axis. The linear relation is demonstrated obviously. By employing the least square fitting, we obtained the correlation between $\Delta R_{np}$ in $^{208}\text{Pb}$ and $L_{0.11}$ as following
\begin{equation}
\Delta R_{np}=(0.0279\pm0.00338)+(0.00307\pm0.00007)L_{0.11}
\end{equation}
where $\Delta R_{np}$ and $L_{0.11}$ are in units of fm and MeV respectively. The gained linear fitting coloured red is fine with the correlation coefficient $r$ up to 0.984. The slope paramter of symmetry energy at the density of $0.11 \text{fm}^{-3}$ $L^{6}_{0.11}$ given by our present work with the finite nuclear properties ranges from 40.5 MeV to 60.3 MeV within the 30 sets of Skyrme interaction. Consequently, the neutron skin thickness $\Delta R_{np} $ of $^{208}\text{Pb}$ can be predicted in terms of this correlation, which is constrained from 0.15 fm to 0.21 fm. If $\Delta R_{np} $ of $^{208}\text{Pb}$ is greater than 0.24 fm, the  direct URCA process to cool down a  $1.4M_{\odot}$ neutron star is allowed \cite{CJ} . Our result  is too small to make the direct Urca process in the $1.4M_{\odot}$ neutron star  occur. There are some recent data about the neutron skin thickness $\Delta R_{np} $ of $^{208}\text{Pb}$, which are shown in Table 2, obtained from various approaches. It appears that our result consists well with  $0.180\pm0.035$ fm from the pygmy dipole resonances (PDR)\cite{AK} and $0.185\pm0.03$ fm from mean field \cite{DZJU}. Recently, Dong et al employed a more practicable strategy than the current lead radius experiment (PREX) to probe the neutron skin thickness of $^{208}\text{Pb}$ based on a high linear correlation between the $\Delta R_{np} $ and $J - a_{\text{sym}}$ \cite{asym}, where $J$ is the symmetry energy (coefficient) of nuclear matter at saturation density. The obtained $\Delta R_{np} $ in $^{208}\text{Pb}$ was $0.176 \pm 0.021$ fm robustly, being consistent with the present result. \\

\setlength{\arrayrulewidth}{0.8pt}
\begin{tablehere}
\caption{The neutron skin thickness $\Delta R_{np} $ of $^{208}\text{Pb}$ probed in various independent studies.} \vspace{1mm}\footnotesize
\begin{center} \doublerulesep 0.1pt \tabcolsep 13.5pt
\begin{tabular}{lccccc}
\hline
Reference&Method&$\Delta R_{\text{np}}$(fm)\\

\hline\
\cite{TA}  & proton inelastic scattering &$0.156^{+0.025}_{-0.021}$\\
\cite{J}& proton elastic scattering&$0.211^{+0.054}_{-0.063}$\\
\cite{KMJ}&chiral effective field theory &$0.17\pm0.03$\\

\cite{DZG}& alpha-decay energies&$0.191\pm 0.032$\\
\cite{DZJU}&mean field    &$0.185 \pm 0.035$\\
\cite{AK}&pygmy dipole resonance&$0.185\pm0.035$\\
\cite{A}&pygmy dipole resonance&$0.194\pm0.024$\\
Present& Nuclear mass differences& 0.15-0.21\\
\hline
\end{tabular}
\end{center}
\end{tablehere}

In addition, we also show the correlations of the neutron skin thickness $\Delta R_{np}$ of  $^{208}\text{Pb}$ and the slope coefficient $L_{0}$ of symmetry energy at the saturation density in the right panel of Fig.1 with a linear fitting. The linear relation is strong with the correlation coefficient $r$=0.964 but is weaker than the one in the left panel. It can be concluded that the neutron skin thickness of heavy nuclei is uniquely fixed by the symmetry energy density slope at a subsaturation density of $0.11 \text{fm}^{-3}$ rather than at the saturation density, which is in agreement with the conclusion in \cite{ZC}.\\
\phantom{fff}After we have got the the slope coefficients $L^{(6)}_{0.11}$ of symmetry energy at the density of $\text{0.11fm}^{-3}$ by 30 sets of Skyrme parameters, with Eq.[2] and Eq.[4], the symmetry energy $S^{(6)}_{0.11}$ and the curvature paramter $K_{\text{sym}}$ at the density of $\text{0.11fm}^{-3}$ can be measured. The calculated $S^{(6)}_{0.11}$ is 24.8-26.3 MeV and $K_{\text{sym}}$ is from -150.1 MeV to -68.3 MeV. Then, the description Eq.[1] for the density-dependent symmetry energy  around the subsaturation density of $\text{0.11fm}^{-3}$ becomes clear. The behaviors of established symmetry energy $S(\rho)$ are presented as a function of density $\rho$ in Fig. 2.\\

\end{multicols}
\begin{figure}[H]
\centering
\includegraphics[scale=0.65]{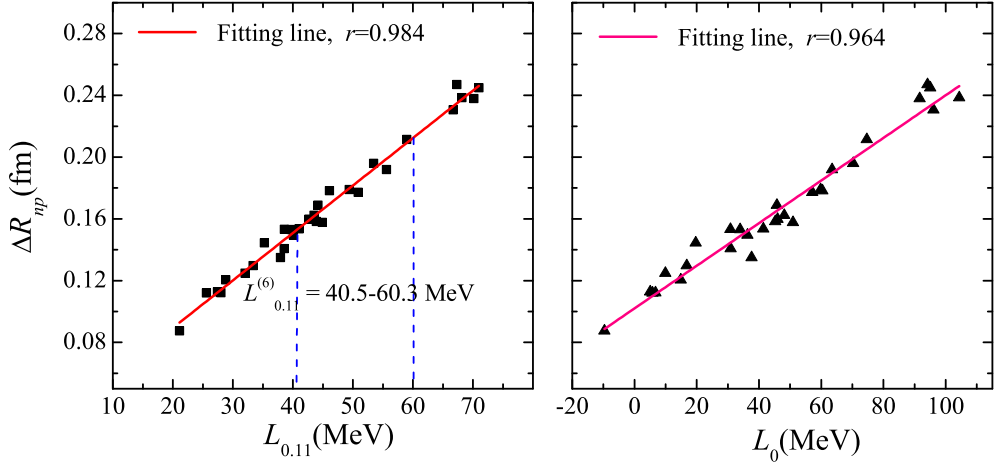}
\caption{Correlation of the neutron skin thickness $\Delta R_{np}$ of  $^{208}\text{Pb}$ and the slope parameter $L_{0.11}$ ($L_{0}$) of symmetry energy at the density of $\text{0.11fm}^{-3}$ (saturation density) in the left (right) panel  within the Skyrme interactions.}  
\label{fig:example2}
\end{figure}
\begin{multicols}{2}

\section{Summary}
In summary, we have employed the symmetry energy coefficient $a_{\text{sym}}(A)$ extracted by experimental nuclear masses  in our previous work and the density profiles of heavy nuclei to explore the behavior of the density-dependent symmetry energy around the subsaturation density of $\text{0.11fm}^{-3}$. The estimated
value of the slope parameter $L_{0.11}$ of symmetry energy at the density of $\text{0.11fm}^{-3}$ is 40.5-60.3 MeV. With the nice linear correlation of $L_{0.11}$ and the neutron skin $\Delta R_{np}$ of  $^{208}\text{Pb}$ , $\Delta R_{np}$ is predicted to be 0.15-0.21 fm, which is too small to allow the direct Urca process in the $1.4M_{\odot}$ neutron stars. It proves that the approach of probing the nuclear matter features by using the information about the nuclear structure is effective.

\begin{figure}[H]
\centering
\includegraphics[scale=0.4]{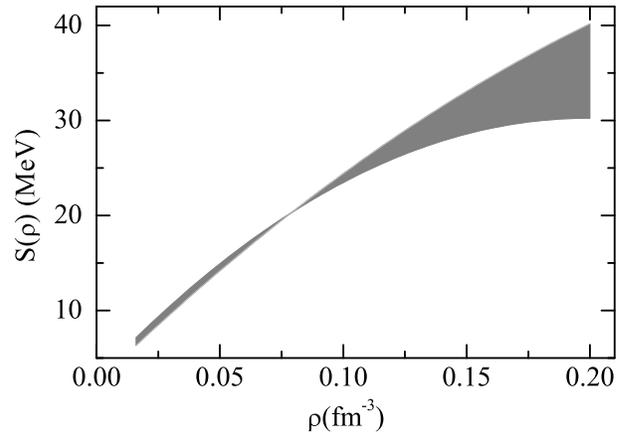}
\caption{Symmetry energy $S(\rho)$ as a function of density $\rho$ with Eq.[5]. } 
\label{fig:example2}
\end{figure}

\vspace*{2mm} \Acknowledgements{\bahao This work was supported by the National Natural Science
Foundation of China under Grants No.
11175219, No. 10975190, No. 11275271 and No. 11405223, by the 973
Program of China under Grant No. 2013CB834405, by the
Knowledge Innovation Project (KJCX2-EW-N01) of Chinese
Academy of Sciences, by the Funds for Creative Research
Groups of China under Grant No. 11321064, and by the Youth
Innovation Promotion Association of Chinese Academy
of Sciences.
}

\end{multicols}


\begin{thebibliography}{99}

\bibitem{FRW}
Danielewicz P, Lacey R, Lynch W G.  Determination of the equation of state of dense matter. science, 2002, 298: 1592-1596
\bibitem{AMJ}
Steiner A W, Prakash M, Lattimer J, et al. Isospin asymmetry in nuclei and neutron stars. Phys Rep, 2005, 411: 325-375
\bibitem{VMV}
Baran V, Colonna M, Greco V, et al. Reaction dynamics with exotic nuclei. Phys Rep, 2005, 410: 335-466
\bibitem{BLC}
Li B A, Chen L W, Ko C M. Recent progress and new challenges in isospin physics with heavy-ion reactions. Phys Rep, 2008, 464: 113-281
\bibitem{ZYX}
Zhang Y X, Danielewicz P, Famiano M, et al. The influence of cluster emission and the symmetry energy on neutron¨Cproton
spectral double ratios. Phys Lett B, 2008, 664: 145-148
\bibitem{S}
Kumar S, Ma Y G, Zhang G Q, et al. Probing the density dependence of the symmetry energy via multifragmentation
at subsaturation densities. Phys Rev C, 2011, 84: 044620
\bibitem{F}
Feng Z Q. Constraining the high-density behavior of the nuclear equation of state from strangeness production in heavy-ion collisions. Phys Rev C, 2011, 83: 067604
\bibitem{Y}
Yong G C. Nuclear symmetry energy and proton-rich reactions at intermediate energies. Phys Rev C, 2011, 84: 014607
\bibitem{M}
Ma W C, Wang F, Ma Y G, et al. Isobaric yield ratios in heavy-ion reactions, and symmetry energy of neutron-rich nuclei at intermediate energies. Phys Rev C, 2011,     83: 064620
\bibitem{G1}
Guo C C, Wang Y J, Li Q F, et al. Influence of the symmetry energy on the balance energy of the directed flow, Sci China-Phys Mech Astron, 2012, 55: 252-259

\bibitem{W3}
Wang Y J, Guo C C, Li Q F, et al. The effect of symmetry potential on the balance energy of light particles emitted from mass symmetric heavy-ion collisions with isotopes, isobars and isotones. Sci China-Phys Mech Astron, 2012, 55: 2407-2413
\bibitem{G}
Gao Y, Yong G C, Wang Y J, et al. Influence of the symmetry energy on the cone-azimuthal emission. Phys Rev C, 2013, 88: 057601
\bibitem{W2}
Wang Y J, Guo C C, Li Q F, et al. Constraining the high- density nuclear symmetry energy with the transverse-momentum dependent elliptic flow. Phys Rev C, 2014 89: 044603
\bibitem{W}
Wu Q H, Zhang Y X, Xiao Z G, et al. Competition between Coulomb and symmetry potential in semi-peripheral heavy ion collisions. Phys Rev C, 2015, 91: 014617

\bibitem{DZ}
Dong J, Zuo W, Scheid W. Correlation between $\alpha$-decay energies of superheavy nuclei involving the effects of symmetry energy. Phys Rev Lett, 2011, 107: 012501

\bibitem{Ren1}
Liu Jian, Ren Zhongzhou, Xu Chang, et al. Systematic study of the symmetry energy under the local density approximation. Phys Rev C, 2013, 88: 024324

\bibitem{Ren2}
Xu Chang, Ren Zhongzhou, Liu Jian. Attempt to link the neutron skin thickness of $^{208}$Pb with the symmetry energy through cluster radioactivity. Phys Rev C, 2014, 90: 064310

\bibitem{CJ}
Horowitz C J, Piekarewicz J. Neutron star structure and the neutron radius of $^{208}\text{Pb}$. Phys Rev Lett, 2001, 86: 5647-5650
\bibitem{BS}
Sharma B K, Pal S. Neutron star structure and the neutron radius of $^{208}\text{Pb}$. Phys Lett B, 2009, 682: 23-26
\bibitem{JM}
Lattimer J M, Prakash M. Nuclear matter and its role in supernovae, neutron stars and compact object binary mergers. Phys Rep, 2004, 333: 121-146; The physics of neutron stars. Science,2004, 304: 536-542
\bibitem{BJ}
Todd-Rutel B G, Piekarewicz J. Neutron-rich nuclei and neutron stars: a new accurately calibrated interaction
for the study of neutron-rich matter. Phys Rev Lett, 2005, 95: 122501
\bibitem{CS}
Horowitz C J, Pollock S J, Souder P A, et al. Parity violating measurements of neutron densities. Phys Rev, 2001, 63: 025501
\bibitem{TM}
Sil T, Centelles M, Vinas X, et al. Atomic parity nonconservation, neutron radii, and effective field theories of nuclei. Phys Rev C, 2005, 71: 045502
\bibitem{DZWZ}
Dong J M,Zhang H F, Wang L J, et al. Density dependence of the symmetry energy probed by $\beta^{-}$-decay energies of odd-A nuclei. Phys Rev C, 2013, 88: 014302

\bibitem{CHEN}
Chen L W. Recent progress on the determination of the symmetry energy. Nucl Phys Rev,  2014, 31: 273-284
\bibitem{B1}
Alex Brown B. Neutron Radii in Nuclei and the Neutron Equation of State. Phys Rev Lett, 2000,  85: 5296-5299
\bibitem{B2}
Typel S, Alex Brown B. Neutron radii and the neutron equation of state in relativistic models. Phys Rev C, 2001, 64: 027302
\bibitem{RJ}
Furnstahl R J. Neutron radii in mean-field models. Nucl Phys A, 2002, 706: 85-110
\bibitem{AW}
Steiner A W, Prakash M, Lattimer J M, et al. Isospin asymmetry in nuclei and neutron stars. Phys Rep, 2005, 411: 325-375
\bibitem{MXXM}
Centelles M, Roca-Maza X, Vinas X, et al. Nuclear symmetry energy probed by neutron skin thickness of nuclei. Phys Rev Lett, 2009, 102: 122502
\bibitem{MXX}
Warda M, Vinas X, Roca-Maza X,  et al. Neutron skin thickness in the droplet model with surface width dependence: Indications of softness of the nuclear symmetry energy. Phys Rev C, 2009, 80: 024316
\bibitem{CJJ}
Horowitz C J, Piekarewicz J. Neutron Star Structure and the Neutron Radius of $^{208}\text{Pb} $. Phys Rev Lett, 2001, 86: 5647-5650
\bibitem{TA}
Tamii A, et al. Complete Electric Dipole Response and the Neutron Skin in $^{208}\text{Pb} $, Phys Rev Lett, 2011, 107: 062502
\bibitem{KMJ}
Hebeler K, Lattimer J M, Pethick C J, et al. Constraints on neutron star radii based on chiral effective field theory interactions. Phys Rev Lett, 2010, 105: 161102

\bibitem{DZG}
Dong J M, Zuo W, Gu J Z. Origin of symmetry energy in finite nuclei and density dependence of nuclear matter symmetry energy from measured $\alpha-decay$ energies. Phys Rev C, 2013, 87: 014303
\bibitem{FD}
Fan X H, Dong J M, Zuo W. Density-dependent nuclear matter symmetry energy at subsaturation densities from nuclear mass differences. Phys Rev C, 2014, 89: 017305
\bibitem{DZJU}
Dong J M, Zuo W, Gu J Z, et al. Density dependence of the nuclear symmetry energy constrained by mean-field calculations. Phys Rev C, 2012, 85: 034308

\bibitem{TD}
Mukhopadhyay T, Basu D N. Nuclear symmetry energy from effective interaction and masses of isospin asymmetric nuclei. Nucl Phys A, 2007, 789: 201-208


\bibitem{SKM}
Samaddar S K, De J N, Vi\~{n}as X, et al. Excitation energy dependence of the symmetry energy of finite nuclei. Phys Rev C, 2007, 76: 041602(R)
\bibitem{MOJA}
Dutra M, Louren\c{c}o O, S\'{a} Martins J S, et al. Skyrme interaction and nuclear matter constraints. Phys Rev C, 2012,  85: 035201
\bibitem{J}
Zenihiro J, Sakaguchi H, Murakami T, et al. Neutron density distributions of $^{204,206,208}\text{Pb}$ deduced via proton elastic scattering at Ep=295 MeV. Phys Rev C, 2010, 82: 044611
\bibitem{AK}
Klimkiewicz A, Paar N, Adrich P, Fallot M, et al. Nuclear symmetry energy and neutron skins derived from pygmy dipole resonances. Phys Rev C, 2007, 76: 051603(R)

\bibitem{A}
Carbone A, Colo G, Bracco A, et al. Constraints on the symmetry energy and neutron skins from pygmy resonances in $^{68}\text{Ni}$ and ${^132}\text{Sn}$. Phys Rev C 2010, 81: 041301(R)

\bibitem{asym}
Jianmin Dong, Wei Zuo, and Jianzhong Gu. Constraints on neutron skin thickness in $^{208}$Pb and density-dependent symmetry energy. Phys Rev C, 2015, 91: 034315

\bibitem{ZC}
Zhang Z, Chen L W. Constraining the symmetry energy at subsaturation densities using isotope binding energy difference and neutron skin thickness. Phys Lett B, 2013, 726: 234-238
\end{thebibliography}
\end{document}